\documentstyle[aps,12pt,preprint,tighten]{revtex}
\begin{document}

\draft

\title{\rightline{{\tt (August 2000)}}
\ \\
Bottom-up model for maximal $\nu_{\mu} - \nu_{\tau}$ mixing}
\author{Nicole F. Bell and Raymond R. Volkas}
\address{School of Physics, Research Centre for High Energy Physics\\
The University of Melbourne, Victoria 3010 Australia\\
(n.bell@physics.unimelb.edu.au, r.volkas@physics.unimelb.edu.au)}
\maketitle

\begin{abstract}
We construct a model which provides maximal mixing between a pseudo-Dirac
$\nu_{\mu}/\nu_{\tau}$ pair, based on a local $U(1)_{L_{\mu}-L_{\tau}}$  symmetry.  Its 
strengths, weaknesses and phenomenological consequences are examined.
The mass gap necessitated by the pseudo-Dirac structure is most naturally
associated with the LSND anomaly. The solar neutrino problem then requires
a light mirror or sterile neutrino. By paying a fine-tuning price to nullify the
mass gap, one can also invoke $\nu_e \to \nu_{\mu,\tau}$ for the solar problem.  
The model predicts a new intermediate range force mediated by the light 
gauge boson of $U(1)_{L_{\mu}-L_{\tau}}$.  Through the mixing of $\mu$, $\tau$
and $e$, this force couples to electrons and thus may be searched for 
in precision ``gravity'' experiments. 
\end{abstract}

\section{Introduction}

Mounting evidence from the SuperKamiokande \cite{atmos,SKatmnu2000} experiment 
suggests that
muon neutrinos are mixed with neutrinos of another flavour, with a mixing angle of
close to $\pi/4$, that is, maximal mixing. In this paper, we adopt the point of
view that the angle $\pi/4$ is a special, unique value that ought to be explained.  
In other words, we will attempt to understand the origin of this maximally large
mixing, as opposed to the view that it is just one possible point in parameter
space which should be assigned no particular significance. The contrast between
this large leptonic mixing angle and the small Cabibbo-Kobayashi-Maskawa mixing
angles in the quark sector is stark, and justifies our point of view.

It is interesting that two-fold maximal mixing can be fairly easily explained if
each active neutrino mixes with a {\it sterile} partner. There are two known ways 
to do
this: embrace the Exact Parity or Mirror Matter Model \cite{mirror}, or suppose a
pseudo-Dirac structure \cite{pseudodirac}. The former seems especially compelling,
because the Exact Parity Model is not much more complicated than the Standard Model
(SM) itself. The pseudo-Dirac structure, while having a degree of elegance in and
of itself, suffers when one requires it to emerge from a complete extension of the
SM. Both possibilities, though, provide a strong theoretical motivation 
for light
sterile neutrino flavours. Also, the combined solar \cite{solar}, atmospheric and 
LSND \cite{lsnd} data
provide interesting indirect experimental support for the existence of at least one
light sterile neutrino.

One of the most important problems in experimental atmospheric neutrino physics at 
present is
to discriminate between the $\nu_{\mu} \to \nu_s$ and $\nu_{\mu} \to \nu_{\tau}$
possibilities. The cleanest atmospheric neutrino data (the fully and partially
contained events) can be explained equally well by both oscillation modes. The
modes can in principle be distinguished by processes sensitive to the matter-effect
(ME) and/or the neutral current (NC). SuperKamiokande has four data sets of this
type: neutrino induced $\pi^0$ production (NC), upward through-going muons (ME),
higher energy partially contained events (ME), 
and multiring events (NC). The $\pi^0$ event sample is
not very useful at present because the production cross-section is poorly known. A
forthcoming measurement of this quantity by the K2K long baseline experiment is
eagerly awaited. SuperKamiokande have recently argued that the last three data sets
disfavour the $\nu_{\mu} \to \nu_s$ scenario \cite{SKatmnu2000}, though this has 
been disputed in Ref.\cite{foots}. 
We await with interest a complete
account of the SuperKamiokande analysis, so that independent researchers can 
judge the robustness
of their conclusion. In any case, the future MINOS and CERN to Gran Sasso long
baseline terrestrial experiments will be able to check whatever conclusions
are drawn on the basis of atmospheric neutrino data.

This paper will be devoted to building a theoretical bottom-up style model for
maximal $\nu_{\mu} - \nu_{\tau}$ mixing. We do so partly to provide a foil for the
mirror and pseudo-Dirac approaches to understanding two-fold maximal mixing. Can
one understand active-active maximal mixing in as compelling a way as one can
active-mirror or active-sterile mixing? In addition, most other proposals for
understanding large angle $\nu_{\mu} - \nu_{\tau}$ mixing \cite{maximal} have invoked grand
unification and/or string motivated physics [such as anomalous U(1) symmetries]. By
contrast, we will use a bottom-up approach, whereby we try to keep the new physics
at as low an energy scale as possible. Also, we will not attempt to connect the
$\nu_{\mu} - \nu_{\tau}$ mixing angle problem with the rest of the flavour (mass
and mixing angle hierarchy) problem. We ask the question: what new physics
principles are implied by the discrete hypothesis of maximal $\nu_{\mu} -
\nu_{\tau}$ mixing? Then, given these general principles, how may they be
instantiated within a complete extension of the SM?

\section{Symmetry principles for maximal $\nu_{\mu} - \nu_{\tau}$ mixing.}

Our experience with the SM strongly suggests that internal symmetry principles play
a very fundamental role in nature. In this section, we will deduce some very simple
symmetry principles suggested by maximal $\nu_{\mu} - \nu_{\tau}$ mixing. Strictly
speaking, the atmospheric neutrino results do not rigorously establish exact
maximal mixing. It is a logical possibility that the mixing is very large but not
maximal. It is reasonable to expect that exact maximal mixing would be correlated
with a symmetry principle, because maximality arises from a special point in parameter
space. For aesthetic reasons, and because of the historical precedent regarding the
importance of symmetries, our fundamental supposition here is that maximal
mixing is our target.

We first make a short and apparently digressive comment. It is interesting to note
that the observation of neutrino oscillations, \footnote{To be precise, neutrino 
{\it oscillations} have yet to be seen - only $\nu_{\mu}$ disappearance has been 
rigorously established.  See, for example, Ref.\cite{lisi}.} 
and hence the necessary introduction
of neutrino masses into the SM, implies degrees of freedom beyond those in the {\it
minimal} SM. This is true irrespective of whether the neutrino masses are of Dirac
or Majorana type. In a sense, therefore, the discovery of neutrino mass is akin to
the previous discoveries of new particles such as the top quark. In another sense,
though, it is dissimilar: the new degrees of freedom implied by neutrino mass
within the gauge theoretic rules of the SM are not uniquely specified, and not {\it
directly} observed (as yet). It is certainly true, however, that renormalisable models
of nonzero neutrino
mass necessitate either an expansion of the fermion sector (right-handed neutrino
states, for instance) or an expansion of the scalar sector (Higgs triplets, for
instance), or both. We will begin with the second possibility, 
by including only the minimal lefthanded neutrino degrees of freedom. In the end,
however,
we will find that there is a natural role within our framework for at
least one light mirror or sterile neutrino state.

Consider a general mass matrix involving $\nu_{\mu L}$ and $\nu_{\tau L}$,
\begin{equation}
\label{mm}
\left[ \begin{array}{cc}
\overline{(\nu_{\mu L})^c} & \overline{(\nu_{\tau L})^c}
\end{array} \right]
\left( \begin{array}{cc}
\delta_{\mu} & m \\ m & \delta_{\tau}
\end{array} \right)
\left( \begin{array}{c}
\nu_{\mu L} \\ \nu_{\tau L}
\end{array} \right),
\end{equation} 
where $\delta_{\mu,\tau}$ are Majorana masses for $\nu_{\mu,\tau}$, while $m$
is a transition mass. The mass eigenvalues are
\begin{equation}
m_{\pm} = \left| \frac{\sqrt{ 4 m^2 + (\delta_{\mu} - \delta_{\tau})^2} \pm (\delta_{\mu}
+ \delta_{\tau})}{2} \right|.
\end{equation}
The mixing angle is given by
\begin{equation}
\tan2\theta = \frac{2m}{\delta_{\tau} - \delta_{\mu}}.
\end{equation}
Very close to maximal mixing arises in the parameter range
\begin{equation}
\text{Case\ 1:}\qquad \delta_{\mu,\tau} \ll m.
\label{pd}
\end{equation}
The mass eigenvalues are then
\begin{equation}
m_{\pm} \simeq m \pm \frac{\delta_{\mu} + \delta_{\tau}}{2},
\end{equation}
with
\begin{equation}
\theta \simeq \frac{\pi}{4} + \frac{\delta_{\mu} - \delta_{\tau}}{4m}.
\end{equation}
This defines a pseudo-Dirac structure \cite{wolfenstein_ps} for the
$\nu_{\mu} - \nu_{\tau}$ system. 
\footnote{This type of active-active 
pseudo-Dirac neutrino is
of course distinct from the active-sterile pseudo-Dirac neutrino states discussed
in the Introduction.}
Alternatively, exact maximal mixing arises if
\begin{equation}
\text{Case\ 2:}\qquad \delta_{\mu} = \delta_{\tau}.
\label{discrete}
\end{equation}
Case 1, with its pseudo-Dirac structure, leads to a nearly degenerate pair of
almost
maximally mixed eigenstates with a mass gap $m$ above zero mass. Case 2
has exact maximal mixing without the necessity of a mass gap. We will see
later that this mass gap is most naturally related to the LSND anomaly.

The symmetry structures underlying the two cases are very simple. Consider Case 1
first. The relatively large transition mass term is invariant under any U(1)
symmetry for which $\nu_{\mu}$ and $\nu_{\tau}$ have opposite charges. The obvious
choice for this symmetry is simply $U(1)_{L_{\mu}-L_{\tau}}$ \cite{lsymm}, 
where $L_{\alpha}$ is
the lepton number for family $\alpha = e,\mu,\tau$. The Majorana mass terms break
this symmetry. The hierarchy $\delta_{\mu,\tau} \ll m$ guarantees, however, that
$U(1)_{L_{\mu}-L_{\tau}}$ is an approximate symmetry correlated with the
pseudo-Dirac structure. As $\delta_{\mu,\tau} \to 0$, the symmetry becomes more
exact and the mixing angle approaches complete maximality (and the masses become
more degenerate). The limiting case of vanishing Majorana masses supplies a
four-component massive neutrino which preserves $L_{\mu} - L_{\tau}$ but breaks
$L_{\mu} + L_{\tau}$ by two units. (The $m$ and $\delta$ terms both break $L_{\mu}
+ L_{\tau}$.) The connection between maximal mixing and increased symmetry
guarantees that the close-to-maximal mixing angle deduced at tree-level will not be
spoiled by radiative corrections (unless some other sector of the theory breaks the
$L_{\mu} - L_{\tau}$ symmetry strongly). This is a well-known property of
pseudo-Dirac states.

Is there any independent reason for considering the $U(1)_{L_{\mu} - L_{\tau}}$
symmetry to be in any way fundamental? Interestingly,
it has been observed \cite{symm,Z'} that $U(1)_{L_{\mu}-L_{\tau}}$ is actually an 
anomaly free symmetry of the minimal (zero neutrino mass) Standard Model and may 
therefore be gauged.
In fact the gauge group of the minimal SM may be enlarged to 
\begin{equation}
SU(3)_c \otimes SU(2)_L \otimes U(1)_Y \otimes U(1)_X
\end{equation}
where $X$ is either $L_{\mu} - L_{\tau}$ or $L_{\tau} - L_{e}$ or  $L_{e} -
L_{\mu}$.
It is important to recognise that it is possible to gauge only one of these three 
alternatives, because anomalies involving two different $X$'s
do not cancel given the minimal SM fermion spectrum. 

Because the pseudo-Dirac structure we want is correlated with the anomaly-free
symmetry $X = L_{\mu} - L_{\tau}$, we shall choose to gauge it, consistent with the
common view that local symmetries are likely to be more fundamental than global
symmetries \cite{witten}. Of course we cannot explain why an $L_{\mu} - L_{\tau}$ 
symmetry should
be given this status, rather than either of the two alternatives. We are simply
suggesting that the maximal mixing between $\nu_{\mu}$ and $\nu_{\tau}$ could be
associated with a gauged $L_{\mu} - L_{\tau}$ which singles out
$\nu_{\mu}$ and $\nu_{\tau}$ as special.

Having identified $L_{\mu} - L_{\tau}$ as playing a crucial role, it is tempting to
speculate about further lines of development. An obvious path is to identify this
quantity with the diagonal generator of a flavour SU(2) symmetry with the second
and third lepton families placed in a doublet. Will not pursue this thought here,
because we want to follow the simplest clues first.

Let us now turn to Case 2. This obviously requires broken $U(1)_{L_{\mu}}$
and $U(1)_{L_{\tau}}$, but the central feature is an unbroken interchange symmetry
$\nu_{\mu L} \leftrightarrow \nu_{\tau L}$ to enforce $\delta_{\mu} =
\delta_{\tau}$. Note that, by contrast to the pseudo-Dirac case, there need be no
hierarchy in the breaking scales for $L_{\mu} + L_{\tau}$ and $L_{\mu} - L_{\tau}$.
The interchange symmetry can arise as a remnant of a fundamental U(2) flavour
symmetry acting on the second and third family of leptons. Observe that SU(2) is
not enough, because the transformation matrix within
\begin{equation}
\left( \begin{array}{c} \nu_{\mu L} \\ \nu_{\tau L} \end{array} \right) \to
\left( \begin{array}{cc} 0 & 1 \\ 1 & 0 \end{array} \right)
\left( \begin{array}{c} \nu_{\mu L} \\ \nu_{\tau L} \end{array} \right)
\end{equation}
is an element of U(2) but not SU(2).

These symmetry principles are simple suggestions for a new physics framework that
could lie behind maximal $\nu_{\mu} - \nu_{\tau}$ mixing. We will now take the
pseudo-Dirac possibility and build a complete extension of the SM around it. We
will not develop Case 2 further in this paper.

\section{Model for a pseudo-Dirac $\nu_{\mu} - \nu_{\tau}$ system.}

\subsection{Basic framework.}

We shall now construct a model which realises the pseudo-Dirac structure 
of Case 1.
Our mass matrix will have the general form of Eq.(\ref{mm}), with each of the 
mass terms arising from the vacuum 
expectation values (VEVs) of $SU(2)_L$ triplet Higgs fields.
Note that the $\delta_{\mu}$ and $\delta_{\tau}$ terms require Higgs field 
having opposite charges under $U(1)_{L_{\mu}-L_{\tau}}$.  For simplicity we shall 
assume that $\delta_{\mu}$ is absent in order to limit the number of Higgs fields.

We wish for the neutrino masses to be naturally tiny, which implies a hierarchy 
between the VEVs of the Higgs triplets and the standard Higgs doublet, which may be 
achieved by invoking the VEV seesaw mechanism. 
This is an appealing scenario whereby the triplets acquire tiny VEVs 
because they have masses much greater than the electroweak scale.

The  Higgs sector we shall consider consists of the following fields, 
\begin{eqnarray}
\phi_0 \sim (1,2,1,0), \;\;\; \chi_0 &\sim& (1,3,2,0), \nonumber  \\
\phi_1 \sim (1,2,1,1), \;\;\; \chi_1 &\sim& (1,3,2,1), \nonumber  \\
\chi_2 &\sim& (1,3,2,2),
\end{eqnarray}
where the numbers label the 
$SU(3)_c \otimes SU(2)_L \otimes U(1)_Y \otimes U(1)_{L_{\mu}-L_{\tau}}$ properties.
Here $\phi_0$ denotes the Standard Model Higgs field.  The additional doublet, 
$\phi_1$, is necessary for the implementation of the VEV seesaw mechanism for 
$\chi_1$ and  $\chi_2$, and will also give rise to off-diagonal terms in the mass 
matrix of the charged fermions.  
The triplet  $\chi_0$, which has no $U(1)_{L_{\mu}-L_{\tau}}$ charge, is responsible 
for the $m$ terms in Eq.(\ref{mm}), while the triplet $\chi_2$, which carries
an $U(1)_{L_{\mu}-L_{\tau}}$ charge of 2, produces the $\delta_{\tau}$ mass term.  
We round out the model with a third triplet $\chi_1$.

Thus the Higgs-fermion couplings are given by
\begin{eqnarray}
{\cal L}^{\text Yuk}_{\nu} &=&
\lambda_{\nu_{\mu\tau}} \overline{\ell}^C_{\mu L} \ell_{\tau L} \chi_0
+ \lambda_{\nu_{\tau}} \overline{\ell}^C_{\tau L} \ell_{\tau L} \chi_2
+\lambda_{\nu_e} \overline{\ell}^C_{e L} \ell_{e L} \chi_0 
+ \lambda_{\nu_{e \tau}} \overline{\ell}_{e L}^C \ell_{\tau L} \chi_1
+ \text{H.c.} \;\;\; \text{and}, \\
{\cal L}^{\text Yuk}_e &=&
\lambda_{\tau e} \overline{\ell}_{e L} \phi_1 \tau_R
+ \lambda_{\mu e} \overline{\ell}_{\mu L} \phi_1 e_R
+ \lambda_e  \overline{\ell}_{e L} \phi_0 e_R
+ \lambda_{\mu}  \overline{\ell}_{\mu L} \phi_0 \mu_R
+ \lambda_{\tau}  \overline{\ell}_{\tau L} \phi_0 \tau_R + \text{H.c},
\end{eqnarray}
where the $\ell$'s are left-handed leptonic doublets.
The Higgs potential is of the form,
\begin{eqnarray}
\label{V}
V &=& \sum_{i} \left[  m_i^2 \phi_i^{\dagger} \phi_i 
+ \frac{1}{2}\lambda_i^2 (\phi_i^{\dagger} \phi_i)^2 \right]
+ \sum_{j} \left[ M_j^2 \chi_j^{\dagger} \chi_j
+\frac{1}{2}\Lambda_j^2 (\chi_j^{\dagger} \chi_j)^2 \right]
\nonumber  \\
&+&\alpha (\phi_0^{\dagger} \phi_0)(\phi_1^{\dagger} \phi_1) 
+\sum_{j \neq j'}\alpha_{jj'} (\chi_j^{\dagger} \chi_j)(\chi_{j'}^{\dagger} \chi_{j'}) 
+ \sum_{ij} \beta_{ij}(\phi_i^{\dagger} \phi_i)(\chi_j^{\dagger} \chi_j) 
\nonumber\\
&+& \left[ \mu_0 \chi_0^{\dagger} \phi_0^2 +\mu_1 \chi_1^{\dagger} \phi_0 \phi_1 
+ \mu_2 \chi_2^{\dagger} \phi_1^2  + \text{H.c.} \right], 
\end{eqnarray}
with $i=0,1$ and $j,j'=0,1,2$.
Denoting the VEVs of the Higgs fields by,
\begin{eqnarray}
\langle \phi_0 \rangle = v_0, \;\;\; \langle \chi_0 \rangle &=& u_0, \nonumber  \\
\langle \phi_1 \rangle = v_1, \;\;\; \langle \chi_1 \rangle &=& u_1 \nonumber  \\
\langle \chi_2 \rangle &=& u_2, 
\end{eqnarray}
it may be observed that for large $M_0,M_1$ and $M_2$, the VEV seesaw relations are 
given by
\begin{eqnarray}
u_0 &\simeq& \frac{\mu_0 v_0^2}{M_0^2},  \nonumber \\
u_1 &\simeq& \frac{\mu_1 v_0 v_1}{M_1^2} \;\;\;\; \text{and} \nonumber \\
u_2 &\simeq& \frac{\mu_2 v_1^2}{M_2^2},
\end{eqnarray}
and thus we may obtain tiny neutrino masses by making $M_0,M_1$ and $M_2$ 
suitably large.

The proliferation of Higgs fields may not be as ad hoc as it appears at first sight.  
Observe that the quantum numbers of the fields are such that
\begin{eqnarray}
\chi_0 &\sim& \phi_0^2 , \nonumber  \\
\chi_1 &\sim& \phi_0 \phi_1, \nonumber  \\
\chi_2 &\sim& \phi_1^2,
\end{eqnarray}
hinting, speculatively, that perhaps the $\chi$'s can be reinterpreted as 
composite objects. 
This would then suggest re-expressing the model in terms of effective operator 
language, making the replacements,
\begin{eqnarray}
\chi_0 &\rightarrow& \frac{1}{M}\phi_0^2 , \nonumber  \\
\chi_1 &\rightarrow& \frac{1}{M}\phi_0 \phi_1, \nonumber  \\
\chi_2 &\rightarrow& \frac{1}{M}\phi_1^2,
\end{eqnarray}
where $M$ is a large mass scale, which of course is connected to the VEV seesaw 
mechanism. 

While the Higgs potential (\ref{V}) is undeniably ugly, it is also true that as
long as the electroweak Higgs particle remains undiscovered, we cannot claim 
to really understand gauge symmetry breaking.  One may wistfully speculate that 
symmetry breaking is actually achieved by a more economical mechanism that we shall 
eventually uncover, and all the Higgs messiness will be re-expressed in more 
elegant language. In the meantime though, we are forced to work with Higgs fields.

In order to realise the pseudo-Dirac form for the $\nu_{\mu}-\nu_{\tau}$ sector, we 
require the hierarchy
\begin{equation}
\label{hier}
u_2 \ll u_0,
\end{equation}
which may be achieved by appropriately adjusting the values of the $M_i$.  
According to our symmetry argument, we would also tend to expect $v_1 < v_0$, 
that is, the VEVs which break $L_{\mu}-L_{\tau}$ ought to be smaller 
than their $L_{\mu}-L_{\tau}$ conserving counterparts. This would also help to 
achieve the hierarchy (\ref{hier}), though it is not essential.
Note that we also expect $u_1 < u_0$.

We may now write down the neutrino mass matrix,
\begin{eqnarray}
\left( \begin{array}{ccc}  \overline{(\nu_{e' L})^C}  &\overline{(\nu_{\mu' L})^C}  
& \overline{(\nu_{\tau' L})^C} \end{array} \right)  
\left( \begin{array}{ccc}
\lambda_{\nu_e}u_0 & 0 & \lambda_{\nu_{e \tau}}u_1  \\ 
0 & 0 & \lambda_{\nu_{\mu\tau}}u_0 \\ 
\lambda_{\nu_{e \tau}}u_1  & \lambda_{\nu_{\mu\tau}}u_0 &  \lambda_{\nu_{\tau}}u_2
\end{array} \right) 
\left( \begin{array}{c} \nu_{e' L} \\ \nu_{\mu' L} \\ \nu_{\tau' L} \end{array} \right),
\end{eqnarray}
where the primes signify that we are not in the basis where the charged lepton mass 
matrix is diagonal.
To first order in $u_1$ and $u_2$ the neutrino masses are
\begin{eqnarray}
m_1 &=& | \lambda_{\nu_e}u_0 |, \nonumber\\
m_2 &=& |-\lambda_{\nu_{\mu\tau}}u_0 +\frac{1}{2} \lambda_{\nu_{\tau}}u_2| 
\;\; \text{and}, \nonumber\\
m_3 &=& |\lambda_{\nu_{\mu\tau}}u_0 + \frac{1}{2} \lambda_{\nu_{\tau}}u_2 |.
\end{eqnarray}
Typical values for the combination of Yukawa coupling constants and the 
triplet VEV's might be
\begin{eqnarray}
|\lambda_{\nu_{\mu\tau}}u_0| &\sim& |\lambda_{\nu_{e}}u_0| 
\sim 1 \text{eV}, \;\; \text{and}  \nonumber\\
|\lambda_{\nu_{\tau}}u_2| & \sim & 10^{-3}-10^{-2} \text{eV}.
\label{typical}
\end{eqnarray}
Nearly maximal mixing between $\nu_{\mu}$ and $\nu_{\tau}$ is guaranteed by the
hierarchy $u_2 \ll u_0$ (provided that the relavant Yukawa coupling
constants do not have a nullifying hierarchy). Mixing between $\nu_e$
and the $\nu_{\mu,\tau}$ system is controlled by $\lambda_{\nu_{e\tau}} u_1$.
Its magnitude will be discussed shortly.

The mass matrix for the charged leptons has the form 
\begin{eqnarray}
\left( \begin{array}{ccc}  
\overline{e'_L}  & \overline{\mu'_L} & \overline{\tau'_L} \end{array} \right)
\left( \begin{array}{ccc}
A  & 0  & E \\ 
D & B & 0 \\ 
0  & 0 & C
\end{array} \right)
\left( \begin{array}{c} e'_R \\ \mu'_R \\ \tau'_R \end{array} \right)
\end{eqnarray}
where $A,B,C \propto v_1$ and $D,E \propto v_2$.

In order to obtain the necessary hierarchy in the fermion masses, we must assume 
some hierarchy in the mass matrix parameters $A, B, C, D$ and $E$.  
Although the choice is by no means unique the possibility which is most natural 
and compelling is $D,E \ll A \ll B \ll C$.  
This is the case where the (off-diagonal) $L_{\mu}-L_{\tau}$ violating 
terms $D$ and $E$, are much smaller than the (diagonal) $L_{\mu}-L_{\tau}$ 
conserving terms $A$, $B$, and $C$. 
In this limit, the charged lepton masses are given by
\begin{eqnarray}
m_e^2 &=& A^2 + O(D^2,E^2) , \nonumber \\
m_{\mu}^2 &=& B^2 + O(D^2,E^2), \nonumber \\
m_{\tau}^2 &=& C^2 + O(D^2,E^2).
\end{eqnarray}
Recall from the Introduction that our ambitions in this paper are
rather limited: we want to address the $\nu_{\mu,\tau}$ maximal mixing
problem without simultaneously solving the entire flavour problem.
So, we just have to live with imposed hierarchies like the above.

\subsection{Without LSND.}

The model building process now presents us with a choice. In this subsection, we
will suppose that the LSND anomaly is not due to neutrino oscillations. 
Degree of freedom economy then suggests that the solar neutrino problem should
be solved by $\nu_e \to \nu_{\mu,\tau}$ oscillations. Let us see what
parameter range will allow this.

We are immediately faced with a difficulty. The electron neutrino mass 
$\sim \lambda_{\nu_e} u_0$ is expected to be of the same order of magnitude
as the average $\nu_{\mu,\tau}$ mass $\sim \lambda_{\nu_{\mu\tau}} u_0$.
The combined effect of the relatively small solar neutrino $\delta m^2$ and the
mass gap arising from the pseudo-Dirac structure is to demand 
a near degeneracy between $\nu_e$ and $\nu_{\mu,\tau}$. This entails some
fine-tuning.

The mass squared 
difference corresponding to the atmospheric neutrino anomaly is
\begin{equation}
\delta m^2_{\text{atmos}} \sim u_0u_2,
\end{equation}
whereas the mass squared difference between the electron neutrino and either of the 
other two mass eigenstates is naturally
\begin{equation}
\delta m^2_{\text{solar}} \sim u_0^2.
\end{equation}
Since $u_2 \ll u_0$, we have to adjust 
$|\lambda_{\nu_e}/\lambda_{\nu_{\mu\tau}}|\simeq 1$ so that,
\begin{equation}
\delta m^2_{\text{solar}} < \delta m^2_{\text{atm}}.
\end{equation}
For example, for either the small or large angle MSW \cite{msw} solutions 
we need $\delta m^2_{\text{solar}} \sim 10^{-5} \text{eV}^2$
which requires $(\lambda_{\nu_e}/\lambda_{\nu_{\mu\tau}})^2$ to be   
fine-tuned to one part in $10^5$ if $m_{2,3} \sim 1$ eV.
This is regrettable, although perhaps not egregiously bad.

The leptonic mixing matrix is given by,
\begin{eqnarray}
U_{\alpha i} \equiv U_e^{\dagger} U_{\nu} 
\simeq 
\left( \begin{array}{ccc}
1  & \epsilon  &\epsilon' \\ 
\epsilon & 1 & 0 \\ 
\epsilon  & 0 & 1
\end{array} \right)
\left( \begin{array}{ccc}
\frac{1}{\sqrt{N_1}}  & \frac{\gamma_1}{\sqrt{N_2}} & \frac{\gamma_2}{\sqrt{N_3}} \\ 
\frac{\gamma_3}{\sqrt{N_1}} & \frac{1}{\sqrt{N_2}}(\frac{1}{\sqrt{2}}+\delta)  
& \frac{1}{\sqrt{N_3}}(\frac{1}{\sqrt{2}}-\delta) \\ 
\frac{\gamma_4}{\sqrt{N_1}} & -\frac{1}{\sqrt{N_2}}(\frac{1}{\sqrt{2}}-\delta) 
& \frac{1}{\sqrt{N_3}}(\frac{1}{\sqrt{2}}+\delta)
\end{array} \right) + O(u_1^2,u_2^2),
\label{leptonicU}
\end{eqnarray}
where 
\begin{equation}
\epsilon = \frac{AD}{B^2} < \left( \frac{m_e}{m_{\mu}} \right)^2, \;\;\;
\epsilon' = \frac{E}{C} < \frac{m_e}{m_{\tau}},
\end{equation}
\begin{eqnarray}
\delta &=& \frac{\lambda_{\nu_{\tau}}}{4\sqrt{2}\lambda_{\nu_{\mu\tau}}}
\frac{u_2}{u_0} \nonumber\\
\gamma_1 &=& \frac{\lambda_{\nu_{e\tau}}}{\sqrt{2}(\lambda_{\nu_{\mu\tau}}
+\lambda_{\nu_e})} \frac{u_1}{u_0} 
\simeq \frac{\lambda_{\nu_{e\tau}}(\lambda_{\nu_{\mu\tau}}-\lambda_e)}
{\sqrt{2}\lambda_{\nu_{\mu\tau}}\lambda_{\nu_{\tau}}}\frac{u_1}{u_2}, \nonumber\\
\gamma_2 &=& \frac{\lambda_{\nu_{e\tau}}}{\sqrt{2}(\lambda_{\nu_{\mu\tau}}-
\lambda_{\nu_e})} \frac{u_1}{u_0}
\simeq \frac{\lambda_{\nu_{e\tau}}(\lambda_{\nu_{\mu\tau}}+\lambda_e)}{\sqrt{2}\lambda_{\nu_{\mu\tau}}\lambda_{\nu_{\tau}}}\frac{u_1}{u_2}, \nonumber\\
\gamma_3 &=& \frac{-\lambda_{\nu_{\mu\tau}}\lambda_{\nu_{e\tau}}}
{(\lambda_{\nu_{\mu\tau}}^2-\lambda^2_{\nu_e})} \frac{u_1}{u_0}
\simeq \frac{-\lambda_{\nu_{e\tau}}}{\lambda_{\nu_{\tau}}}\frac{u_1}{u_2}, 
\nonumber\\
\gamma_4 &=& \frac{-\lambda_{\nu_e} \lambda_{\nu_{e\tau}}}
{(\lambda_{\nu_{\mu\tau}}^2-\lambda^2_{\nu_e})} \frac{u_1}{u_0}
\simeq -\left( \frac{\lambda_{\nu_e}}{\lambda_{\nu_{\mu\tau}}} \right)\frac{\lambda_{\nu_{e\tau}}}{\lambda_{\nu_{\tau}}}\frac{u_1}{u_2},
\label{gamma}
\end{eqnarray}
and
\begin{eqnarray}
N_1 &\simeq& 1 + \gamma_3^2 + \gamma_4^2 \nonumber\\
N_2 &\simeq& 1 + \gamma_1 ^2 + O(\delta^2) \nonumber\\
N_3 &\simeq& 1 + \gamma_2 ^2 + O(\delta^2)
\end{eqnarray}

The second set of near equalities in eq.(\ref{gamma}) is related to 
the fine-tuning 
$|\lambda_{\nu_e}/\lambda_{\nu_{\mu\tau}}| \simeq 1 $, or more specifically,
$(\lambda_{\nu_{\mu\tau}}^2-\lambda^2_{\nu_e})u_0^2 \simeq \lambda_{\nu_{\tau}} 
\lambda_{\nu_{\mu\tau}}u_0u_2$.

The values of $\gamma_i$ could be such as to provide either a small or large 
angle solar MSW solution.
\footnote{We note that recent Superkamiokande data disfavour the small-angle MSW solution
\cite{SKsolarnu2000}.  The statistical significance is, however, not yet severe
enough to make this type of solution completely uninteresting.}
For example, with 
$\lambda_{\nu_{e\tau}}u_1 \simeq 0.04 \lambda_{\nu_{\tau}}u_2$, we would 
have the small angle solution with 
$\sin \theta_{\text{solar}}\simeq 0.05$.
Alternatively, consider for example 
$\lambda_{\nu_e} \simeq - \lambda_{\nu_{\mu\tau}}$, and 
$\lambda_{\nu_{e\tau}}u_1 \simeq 0.3 \lambda_{\nu_{\tau}}u_2$.  We then obtain 
a large angle solution with 
$\sin \theta_{\text{solar}} \simeq \gamma_1 \simeq 0.4$, and $\gamma_2 \ll 1$.

\subsection{With LSND.}

As we have just seen, the pseudo-Dirac mass gap causes some problems  with
solving the solar neutrino problem. A more elegant and natural
alternative is to exploit the mass gap, rather than trying to fight it.
The $\nu_e$ and $\nu_{\mu,\tau}$ mass eigenvalues are naturally of the
same order, with the scale set by $u_0$. The $\nu_e - \nu_{\mu,\tau}$
$\delta m^2$ scale is thus of order $u_0^2$ if we do not fine tune a near
degeneracy.

Indeed, the guideline values of eq.(\ref{typical}) put the $\nu_e - \nu_{\mu}$
$\delta m^2$ scale in the LSND range. It is certainly noteworthy that the
LSND scale is a few orders of magnitude larger than the solar and atmospheric
scales, and it is quite attractive to associate this higher scale with the
mass gap {\it which was constructed for another reason}.

The LSND mixing angle must come out of the first equalities in eq.(\ref{gamma})
(the near equalities do not hold in the absence of the previous
fine tuning). Using eq.(\ref{typical}) as a guide, we see that 
$\lambda_{\nu_{e\tau}} u_1$ should be of order $0.1$ eV to put $\gamma_1$
or $\gamma_2$ in the LSND mixing angle range. Since this is intermediate
between the $u_0$ and $u_2$ scales, it fits in nicely with our
fundamental $u_2 < u_1 < u_0$ symmetry breaking pattern, dictated by
approximate $U(1)_{L_{\mu} - L_{\tau}}$ symmetry.

The solar neutrino problem then requires the introduction of a light
sterile neutrino $\nu_s$ which mixes with $\nu_e$. The most attractive
possibility is to have this mixing also being maximal. This scenario
fits all of the data, except for Homestake, extremely well \cite{mirror,max}. 
The proper
incorporation of a light sterile neutrino into the model is really
beyond the scope of this paper. However, it is pretty obvious that
the mirror matter or exact parity idea \cite{mirror} is quite relevant, not only
for providing a reason for the sterile state to be light, but also
to explain the $\nu_e/\nu_s$ maximal mixing. One could imagine that
the mirror matter solution to the neutrino anomalies summarised
in Ref.\cite{mirror} is half correct: that the solar oscillations are into
mirror partners, but the atmospheric oscillations are $\nu_{\mu}
\leftrightarrow \nu_{\tau}$. Intriguingly, Ref.\cite{yoon} has also recently
canvassed this possibility.

\section{Other phenomenology.}

Let us now consider constraints on the model.  There will be a new gauge 
boson $Z'$, corresponding to the $U(1)_{L_{\mu}-L_{\tau}}$ gauge symmetry.
Due to the VEV of $\phi_1$, there will be a mass mixing  term involving the $Z'$ 
and the ordinary $Z$ gauge boson.  In order not to significantly modify the 
properties of the standard Z boson, the $Z'$ boson must have a mass that is
much smaller.

The constraints on the model, however, are not terribly stringent, since the 
family lepton number violating processes involve essentially only the neutrinos.  
We suppress the processes involving the charged leptons simply by making the 
off-diagonal ($L_{\mu}-L_{\tau}$ violating) terms $D$ and $E$ in the mass 
matrix suitably 
small, which will also avoid flavour changing neutral current type processes due 
to exchange of $\phi  $ bosons. 
Note that this is not an ad hoc requirement, but is in perfect accord with 
our symmetry argument.

There is also the issue of kinetic mixing of the neutral gauge bosons to consider, 
which arises whenever you have a theory with two local $U(1)$ symmetries.
The kinetic part of the Lagrangian may be written as
\begin{equation}
{\cal L}^{\text{kinetic}} = -\frac{1}{4}F^{\mu\nu}F_{\mu\nu} 
- \frac{1}{4}F'^{\mu\nu}F'_{\mu\nu} -\frac{2}{4} \kappa F^{\mu\nu}F'_{\mu\nu},
\end{equation} 
where, in general, the kinetic mixing term, $\kappa$, will arise directly in the 
Lagrangian \cite{robert}, though it may also be generated radiatively \cite{holdom}.
The physical neutral gauge boson states are then found by diagonalising the 
$Z$ and $Z'$ kinetic and mass terms \cite{robert}. This alters the coupling of the 
$L_{\mu}-L_{\tau}$ gauge boson by adding a term proportional to the hypercharge - 
hence $Z'$ will couple not only to the leptons but also to the quarks, though in a 
generation independent fashion. We have to assume that $\kappa$ is small.

An interesting point to consider, since the $Z'$ gauge boson is light, is whether there 
will be any ``5th force'' effects.  
In other words, an effective violation of the equivalence principle through 
a tiny $Z'$ boson mediated repulsion of matter \cite{price,adelberger}.
In fact, for suitable parameters, a signature of the model would 
be a new intermediate range force of nature.
The $Z,Z'$ mass matrix is
\begin{eqnarray}
\left( \begin{array}{cc}
\frac{g^2}{8 \cos^2 \theta_W}(v_0^2+v_1^2) & \frac{gg'}{8 \cos \theta_W} v_1^2  \\
\frac{gg'}{8 \cos \theta_W}v_1^2 & \frac{g'^2}{8} v_1^2  
\end{array} \right),
\end{eqnarray}
where $g'$ denotes the gauge coupling constant of the 
$U(1)_{L_{\mu}-L_{\tau}}$ symmetry, and for simplicity, kinetic mixing 
and the tiny triplet VEVs have been neglected.
The light eigenstate, which will be predominately the $Z'$, will have a mass given by
\begin{equation}
M^2_{Z_{\text{light}}} \simeq \frac{g'^2}{8} \frac{v_1^2v_0^2}{v_1^2 + v_0^2}.
\end{equation}
Of course, in order for the $Z'$ to be detectable through violation of the 
equivalence principle its mass must be incredibly tiny.  For example, if the mass 
of the $Z'$ was larger than say $10^{-5}$ eV, the corresponding range of the force 
would be less than of order 1cm.
The electron couples to the light gauge boson, both through mixing between the 
$e,\mu$ and $\tau$, and mixing of $Z$ and $Z'$. 
The coupling is given by
\begin{eqnarray}
\label{coupl}
&&\frac{1}{2}g'\cos\theta_Z(\epsilon^2-\epsilon'^2)\overline{e}_L Z\!\!\!\!/_{\text{light}}e_L
+\frac{1}{2}g'\cos\theta_Z(\zeta^2-\zeta'^2)\overline{e}_R Z\!\!\!\!/_{\text{light}}e_R 
\nonumber\\
&& \;\;\; + \sin \theta_Z \frac{g}{4\cos \theta_W} \overline{e} Z\!\!\!\!/_{\text{light}} 
(1-4\sin\theta_W-\gamma_5)e
\end{eqnarray}
where $\theta_Z$ is the $Z,Z'$ mixing angle such that
\begin{equation}
\sin\theta_Z \simeq \tan\theta_Z = \frac{g'\cos\theta_W}{g} \frac{v_1^2}{v_0^2+v_1^2},
\end{equation}
and
\begin{equation}
\zeta = \frac{D}{B}, \;\;\;
\zeta' = \frac{AE}{C^2}.
\end{equation}
Note that
the strength of the force is diminished both by $g'$ and the small parameters
$\epsilon^2$, $\epsilon'^2$, $\zeta^2$, $\zeta'^2$ and $(v_1/v_0)^2$. If one demands
consistency with standard Big Bang Nucleosynthesis, then the very rough bound
$g' \stackrel{<}{\sim} 10^{-10}$ is indicated to prevent the $Z'$ boson being
in thermal equilibrium during the relevant epoch of the early universe.

Tests of the equivalence principle on short distance scales are the subject of 
a proposed experiment \cite{price} which aims to explore the range from about 
$10\mu{\rm m}$ to $1{\rm cm}$, for forces of strength relative to gravity of
about $10^{-2}$ upward.  See Fig.1 of Ref.\cite{price}.  
The predicted $Z'$ gauge boson may be observable 
in this experiment, for a certain parameter region. 

The Yukawa potential associated with the $Z'$ gauge boson will be of the form
\begin{equation}
V_{{\rm Yukawa}} = f^2 \frac{e^{-r/\lambda}}{r},
\end{equation} 
where $\lambda \simeq 1/M_{{\rm Z_{light}}}$ is the range of the force.
For the coupling to electrons given by Eq.(\ref{coupl}) we have
\begin{equation}
f^2/\hbar c \sim g'^2 \left( \frac{v_1}{v_0} \right)^4,
\end{equation}
and the strength of the force, relative to gravity, will be 
\begin{equation}
\alpha = \frac{f^2}{G_N u^2},
\end{equation}
where $u$ is the atomic mass, and $G_N$ is the Newtonian gravitational constant.
The range and strength of the force have the approximate values
\begin{eqnarray}
\lambda &\sim& 10^{-8}\left( \frac{v_0}{v_1} \right) \left( \frac{10^{-10}}{g'} \right) {\rm meters}, \nonumber \\
\alpha &\sim& 10^{18}\left( \frac{v_1}{v_0} \right)^4 \left( \frac{g'}{10^{-10}} \right)^2. 
\end{eqnarray}
If we assume for example $g' \sim 10^{-10}$ and $ v_1/v_0 \sim 0.1$, the 
range $\lambda$ will be too short for the force to be detected, while for 
smaller values of $v_1/v_0$ it should be observable in the proposed 
experiment \cite{price}.  If we assume smaller values of $g'$, the range 
of the force increases and would violate current experimental constraints.

\section{conclusion}

We have taken a bottom-up approach to constructing a model with maximal mixing between 
two standard neutrinos.  The simple $U(1)_{L_{\mu}-L_{\tau}}$ symmetry naturally provides 
a pseudo-Dirac form for $\nu_{\mu}$ and $\nu_{\tau}$ with close to maximal mixing,
with a mass gap. 
The model can incorporate mixing with the $\nu_e$, consistent 
with either a small or large  angle MSW solution of the solar neutrino anomaly.
However, a mild fine-tuning price must be paid to achieve 
$\delta m^2_{\text{solar}} <\delta m^2_{\text{atmos}}$.  If this price is taken to be 
too high, then one needs to introduce a light mirror or sterile neutrino to solve 
the solar deficit problem by $\nu_e \rightarrow \nu_s$.  It is certainly
suggestive that the natural scale for the $\nu_e-\nu_{\mu}$ mass squared
difference could be in the LSND range, due to the mass gap required by the
pseudo-Dirac structure.
The breaking of the $U(1)_{L_{\mu}-L_{\tau}}$ symmetry occurs at a low scale, and we 
predict a new intermediate range force which may be detectable as an apparent 
violation of the equivalence principle.

\acknowledgments{RRV is supported by the Australian Research Council. 
NFB is supported by the Commonwealth of Australia and The University
of Melbourne. We thank T. L. Yoon and R. Foot for somewhat belated discussions.}

\end{document}